\documentclass[aps,prb,floatfix,amsmath,amssymb,preprint]{revtex4}
\usepackage{graphicx}
\usepackage{dcolumn}
\usepackage{bm}
\usepackage{epstopdf}
\usepackage{enumerate}
\usepackage{setspace}   
\usepackage{amsmath}
\usepackage{amssymb}

\begin{document}
\title{Tensor networks---a new tool for old problems}

\author{Subir Sachdev}
\affiliation{Department of Physics, Harvard University, Cambridge MA
02138}
\begin{abstract}
A new renormalization group approach that maps lattice problems to tensor networks may hold the key to solving seemingly intractable models of strongly correlated systems in any dimension.\\
~\\
\begin{center}
{\em Physics\/} {\bf 2}, 90 (2009)\\
Published October 26, 2009\\
DOI: 10.1103/Physics.2.90\\
URL: http://link.aps.org/doi/10.1103/Physics.2.90
\end{center}
~\\
\\
A Viewpoint on:\\
{\em Tensor-entanglement-filtering renormalization approach and symmetry-protected topological order\/}\\
Zheng-Cheng Gu and Xiao-Gang Wen Phys. Rev. B {\bf 80}, 155131 (2009).
\end{abstract}

\maketitle

Some of the deepest and most difficult problems in mathematics are also often the simplest to state.
The most celebrated example is Fermat's last theorem, whose statement can be understood
by high school students, and yet required the full arsenal of advanced modern mathematics to prove.

In condensed matter physics, we are also faced with a difficult class of  problems which can be stated quite simply,
using material familiar to every beginning student of quantum mechanics. There has been a sustained effort
to attack these problems by numerous physicists for over two decades, 
but there is only partial progress to report. The recent paper
{\em Tensor-Entanglement-Filtering Renormalization Approach
and Symmetry Protected Topological Order}, by Zheng-Cheng Gu and Xiao-Gang Wen, 
adds another promising tool with which such problems can be
addressed. However, it remains to be seen if it will finally break the logjam and lead to a comprehensive solution.

The simplest of these problems involve only the spin operators ${\bf S}_i$ of electrons residing on the
sites, $i$, of a regular lattice. Each electron can have its spin oriented either up or down, leading to a Hilbert
space of $2^N$ states, on a lattice of $N$ sites. On this space acts the Hamiltonian
\begin{equation}
\mathcal{H} = \sum_{i < j} J_{ij} {\bf S}_i \cdot {\bf S}_j
\end{equation}
where the $J_{ij}$ are a set of short-range exchange interactions, the strongest of which have $J_{ij} > 0$ {\em i.e.\/}
are antiferromagnetic. We would like to map the ground
state phase diagram of $\mathcal{H}$ as a function of the $J_{ij}$ for a variety of lattices in the limit of $N \rightarrow \infty$.
Note that we are not interested in obtaining the exact wavefunction of the ground state: this is a hopeless task
in dimensions greater than one. Rather we would be satisfied in a qualitative characterization of each phase in the space
of the $J_{ij}$.
Among possible phases are:\\ 
({\em i\/}) a N\'eel phase, in which the spins have a definite orientation just as in the classical antiferromagnet, with 
the $\langle {\bf S}_i \rangle$ all parallel or anti-parallel to each other.\\
({\em ii\/}) a spiral antiferromagnet, which is magnetically ordered like the N\'eel phase, but the $\langle {\bf S}_i \rangle$
are not collinear,\\
({\em iii\/}) a valence bond solid (VBS), with the spins paired into $S=0$ valence bonds which crystallize into a 
preferred arrangement which breaks the lattice symmetry, and\\
({\em iv\/}) a spin liquid, with no broken symmetries, neutral $S=1/2$ elementary excitations, and varieties of a subtle
`topological' order.

For a certain class of $\mathcal{H}$, the above problem has effectively been solved using fast computers.
These are lattices for which the Feynman path integral for $\mathcal{H}$ can be evaluated as a sum over
configurations with positive weights; the sum is then evaluated by sampling based upon the Monte Carlo method.
A prominent example of this solution is the recent work by Lou {\em et al.} \cite{sandvik} on a set of square lattice antiferromagnets, in which
they find N\'eel and VBS states.

\begin{figure}
\begin{center}
\includegraphics[width=3in]{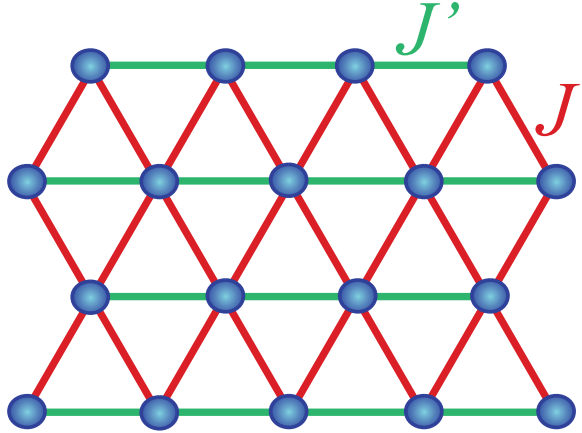}
\caption{Distorted triangular lattice representing the geometry of systems studed experimentally
in Ref.~\onlinecite{kato6}.} \label{lattice}
\end{center}
\end{figure}

However, there are a large class of lattices for which the path integral does not have any known representation
with only positive weights. The Monte Carlo method cannot be used here - this is the famous {\em sign \/} problem.
An important example is the model on the triangular lattice, with nearest neighbor couplings $J$ and $J'$ as illustrated
in Fig.~\ref{lattice}. 
This model is of experimental importance: the organic insulators X[Pd(dmit)$_2$]$_2$ are modeled
by a range of values of $J'/J$ as X is varied over a series of monovalent cations \cite{kato6}, and N\'eel, VBS, and a candidate spin liquid
phase have been discovered.

The sign problem has effectively been conquered in one dimension, by the density matrix renormalization group \cite{dmrg} (DMRG).
Its success has spawned an intense effort to discover a generalization which works in two and higher dimensions.
In recent years, ideas from quantum information theory have been particularly influential: the ground states
of models like $\mathcal{H}$ have subsystems with an entanglement entropy which scales with the boundary area, and methods have been devised
which restrict the numerical sampling to only such states. There is an alphabet soup of proposals \cite{cirac}, including matrix products states (MPS),
projected entangled-pair states (PEPS), multi-scale renormalization
ansatz \cite{vidal} (MERA), tensor renormalization group \cite{levin} (TRG), 
and now the tensor-entanglement-filtering renormalization (TEFR) of Gu and Wen. These methods are connected to each other,
and differ mainly in the numerical algorithm used to explore the possible states.
So far no previously unsolved model $\mathcal{H}$ has been moved into the solved column, although recent results from Evenbly and
Vidal \cite{kagome}
show fairly conclusive evidence for VBS order on the kagome lattice, and there is promising progress on 
frustrated square lattice antiferromagnets \cite{ciracj1j2j3}.

\begin{figure}
\begin{center}
\includegraphics[width=3.9in]{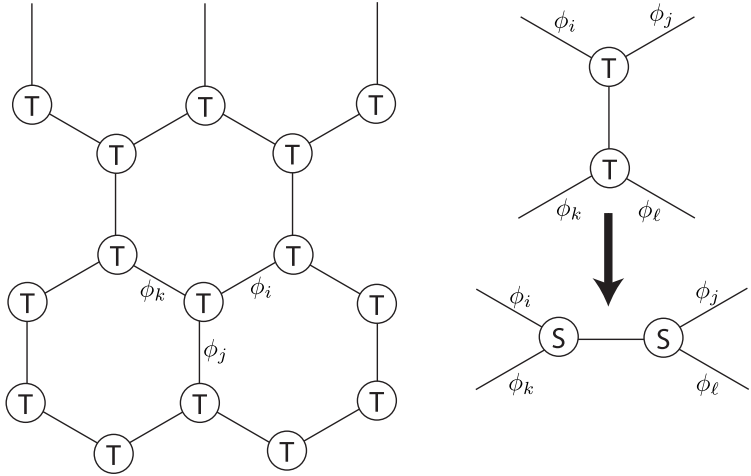}
\caption{Left: representation of the partition function as a trace over the indices of the third-rank tensors $T$, with
their indices contracted along the links of the lattice. Right: A step in the TRG replacing $T$ by a new tensor $S$.} \label{tensor}
\end{center}
\end{figure}

The TEFR descends from the TRG of Levin and Nave \cite{levin}. They consider a rewriting of the spacetime partition function
of $\mathcal{H}$ in terms of a discrete field $\phi_i$, which resides on the {\em links\/} of a spacetime lattice (not necessarily the same
lattice as that of $\mathcal{H}$). Then, for a very general class of $\mathcal{H}$ with local interactions, the partition function
can be written as
\begin{equation}
\mathcal{Z} = \sum_{\phi_i, \phi_j, \phi_k, \ldots} \prod T_{\phi_i, \phi_j, \phi_k}
\label{mathcalz}
\end{equation}
where $T$ is a tensor on the sites of a spacetime lattice whose components are labeled by the allowed values
of the $\phi_i$. This construction is illustrated in Fig.~\ref{tensor} for the honeycomb lattice, in which case
$T$ is a third-rank tensor, as is assumed in Eq.~(\ref{mathcalz}). 
Note that the sum over $\phi_i$ corresponds to a contraction
of the tensor indices across each link of the lattice, and $\mathcal{Z}$ has all indices contracted (for periodic boundary conditions).

The key step in the TRG is a coarse-graining of Eq.~(\ref{mathcalz}) to a more dilute lattice, as in any renormalization group (RG)
transformation. In the conventional Wilsonian real-space RG, this is done by just
summing over a select subset of the $\phi_i$. However, the TRG is defined in a way which preserves the virtues of the DMRG in efficiently preserving
the local connectivity information for a variety of neighborhood environments. The important step is the transformation illustrated
in Fig.~\ref{tensor} in which the tensor $T$ is replaced by a new tensor $S$ with a different local connectivity.

The TEFR is an improvement of the TRG which efficiently removes redundant information on local degrees of freedom which eventually
decouple from the long distance behavior, and are not crucial in characterizing the quantum state. This is done by a set of `disentangling'
operations during the coarse-graining procedure. The benefit is a nearly one-to-one correspondence between the fixed-point values
of the local tensor and the identification of the quantum state. Thus each of the states ({\em i\/})-({\em iv}) would 
correspond to distinct values of the fixed-point tensor. In particular Gu and Wen claim that their method also distinguishes the subtle varieties
of topological order in the different spin liquid states.

So far, Gu and Wen have illustrated their method for one-dimensional quantum systems. In these cases, their results are in excellent
accord with field-theoretic predictions and the results of DMRG. It remains to be seen if they can break the logjam in two and higher
dimensions.


\begin{thebibliography}{99}

\bibitem{sandvik} J.~Lou, A.~W.~Sandvik, and N.~Kawashima, Phys. Rev. B {\bf 80}, 180414 (2009).

\bibitem{kato6}  Y.~Shimizu, H.~Akimoto, H.~Tsujii, A.~Tajima, and R.~Kato, J. Phys.: Condens. Matter {\bf 19}, 145240 (2007).

\bibitem{dmrg} S.~R.~White, Phys. Rev. Lett. {\bf 69}, 2863 (1992).

\bibitem{cirac} J.~I.~Cirac and F.~Verstraete, J. Phys. A: Math. Theor. {\bf 42}, 504004 (2009).

\bibitem{vidal} G.~Vidal, Phys. Rev. Lett. {\bf 99}, 220405 (2007).

\bibitem{levin} M.~Levin and C.~P.~Nave, Phys. Rev. Lett. {\bf 99} 120601 (2007).

\bibitem{kagome} G.~Evenbly and G.~Vidal, Phys. Rev. Lett. {\bf 104}, 187203 (2010)

\bibitem{ciracj1j2j3} V.~Murg, F.~Verstraete, and J.~I.~Cirac, Phys. Rev. B {\bf 79}, 195119 (2009).

\end{thebibliography}
\end{document}